\documentclass[unnumsec,webpdf,modern,large,namedate]{oup-authoring-template}

\usepackage{datetime}

\graphicspath{{Fig/}}

\theoremstyle{thmstyleone}%
\theoremstyle{thmstyletwo}%
\theoremstyle{thmstylethree}%

\usepackage{etoolbox}
\usepackage{xcolor}
\newtoggle{revision}
\newcommand{\revision}[1]{\iftoggle{revision}{{\color{red}#1}}{#1}}

\definecolor{darkblue}{rgb}{0.2, 0.2, 0.6}
\hypersetup{colorlinks,linkcolor=black,citecolor=darkblue,urlcolor=black}

\usepackage[english]{babel}

\begin{document}

\journaltitle{arXiv}
\DOI{10.48550/arXiv.2310.06725}
\copyrightyear{2023}
\pubyear{2023}
\access{Advance Access Publication Date: \today}
\appnotes{Review}

\firstpage{1}


\title[Deep learning methods for protein interactions]{Growing ecosystem of deep learning methods for modeling protein--protein interactions}

\author[1,$\ast$]{Julia R. Rogers\ORCID{0000-0003-3269-6375}}
\author[1]{Gergő Nikolényi\ORCID{0000-0003-0581-7864}}
\author[1,$\ast$]{Mohammed AlQuraishi\ORCID{0000-0001-6817-1322}}

\authormark{Rogers et al.}

\address[1]{\orgdiv{Department of Systems Biology}, \orgname{Columbia University}, \orgaddress{New York, \state{New York}, \country{USA}, \postcode{10032}}}

\corresp[$\ast$]{Corresponding author. \href{email:email-id.com}{m.alquraishi@columbia.edu}; \href{email:email-id.com}{jr4182@cumc.columbia.edu}}



\abstract{Numerous cellular functions rely on protein--protein interactions. Efforts to comprehensively characterize them remain challenged however by the diversity of molecular recognition mechanisms employed within the proteome. Deep learning has emerged as a promising approach for tackling this problem by exploiting both experimental data and basic biophysical knowledge about protein interactions. Here, we review the growing ecosystem of deep learning methods for modeling protein interactions, highlighting the diversity of these biophysically-informed models and their respective trade-offs. We discuss recent successes in using representation learning to capture complex features pertinent to predicting protein interactions and interaction sites, geometric deep learning to reason over protein structures and predict complex structures, and generative modeling to design \textit{de novo} protein assemblies. We also outline some of the outstanding challenges and promising new directions. Opportunities abound to discover novel interactions, elucidate their physical mechanisms, and engineer binders to modulate their functions using deep learning and, ultimately, unravel how protein interactions orchestrate complex cellular behaviors.
}


\maketitle

\section{Introduction}

Protein--protein interactions underpin numerous molecular processes that are essential to cellular functions ranging from metabolism and growth to motility and development. For example, the enzyme responsible for producing chemical energy assembles from two multisubunit protein motors~\citep{Guo2019}; nearly one thousand protein subunits form nuclear pore complexes to regulate traffic of genetic messages into the cytoplasm~\citep{Mosalaganti2021}; and networks of actin filaments provide a structural and mechanical support system that enables cellular motion~\citep{Reynolds2022}. A comprehensive understanding of protein interactomes thus promises fundamental insights into cellular behavior. Such knowledge is key to ultimately deciphering how aberrant protein interactions manifest in disease~\citep{Cheng2021} and engineering protein interaction modulators for therapeutic applications~\citep{Lu2020}.

\begin{figure}[htb!]
    \centering
    \includegraphics[width=213pt]{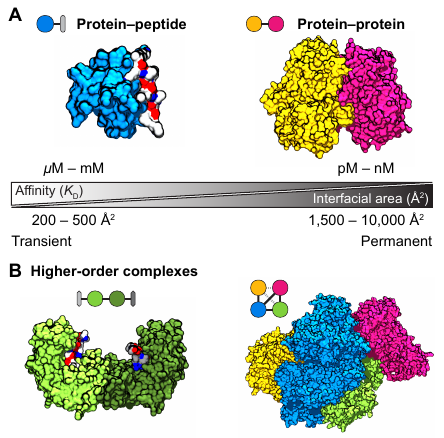}
    \caption{\textbf{Diversity of protein interactions.} (A) Direct, physical interactions involve a protein and peptide (typically defined by lengths of less than 25 residues) or a pair of proteins. Interaction affinity, as quantified by dissociation constant ($K_{\rm D}$), distinguish weak, transient interactions with $K_{\rm D} \ge \mu$M from strong, permanent interactions with $K_{\rm D} \ll \mu{\rm M}$. Interaction surface areas are smallest for protein--peptide interactions and range from $500 - 1,500$~\AA\ for weak protein--protein interactions to $10,000$~\AA\ for obligate ones. (B) Proteins assemble into higher-order complexes through combinations of direct, pairwise interactions. Examples illustrate a homodimeric complex with two direct protein--peptide interactions and a heterooligomeric protein tetramer. Solid black lines indicate direct interactions while dashed gray ones indicate indirect ones where molecular surfaces are not in physical contact.}
    \label{fig:ppi_types}
\end{figure}

Protein interactions coordinate such a wide array of cellular processes by utilizing an equally varied set of molecular recognition mechanisms. For example, very strong---effectively permanent with negligible probability of being unbound---obligate interactions ensure the stability of essential, multisubunit enzymes, whereas weak, transient interactions involving regulatory or signaling proteins allow for rapid responses to environmental stimuli~\citep{AcunerOzbacan2011,Kastritis2013}. These two examples sit at extremes of a continuous spectrum of direct, physical interactions~\citep{Peng2017,Mosca2013_interactome3d} with varied affinities, interfacial sizes, and structural properties (Fig.~\ref{fig:ppi_types}A).
Strong protein interactions often exhibit large, hydrophobic interaction surfaces, whereas transient ones display a larger fraction of polar and charged residues that can favorably interact with solvent molecules in their unbound states~\citep{AcunerOzbacan2011,Kastritis2013}. Interactions mediated by short peptidic motifs of roughly ten contiguous residues exemplify the weak end of the spectrum (with affinities in the micromolar range) employed to dynamically control protein localization, activation, and degradation~\citep{VanRoey2014}. Often a few interfacial `hot spot' residues contribute substantially to interaction affinity and specificity; however, residues distant from the interaction surface can also impact binding, for instance by altering unbound monomers' conformational ensembles~\citep{Kastritis2013}. Indeed, some proteins undergo large conformational changes upon binding, and many peptidic sites found within intrinsically disordered regions only adopt well-defined structures upon binding to a receptor~\citep{VanRoey2014}. Yet other proteins maintain `near-rigid' structures, exhibiting only minor fluctuations in side chain orientations upon binding~\citep{Kastritis2013}. One example is the subpicomolar-affinity binding of protease inhibitors to their cognate enzymes~\citep{Kastritis2013}, which is well approximated by a lock-and-key model.

To catalogue and characterize the full span of pairwise interactions that organize proteins into functional complexes (Fig.~\ref{fig:ppi_types}), multiple approaches, with varying trade-offs, have been employed. High-throughput, qualitative experimental assays enable large-scale mapping of protein interaction networks~\citep{Luck2020,Huttlin2021} at reduced sensitivity and specificity relative to low-throughput methods~\citep{AcunerOzbacan2011,Davey2023}. Nevertheless, these interaction maps remain sparse drafts---for example they are estimated to encompass only $2-11$\% of all pairwise interactions in humans~\citep{Luck2020}---and biased towards stable interactions involving high-abundance proteins in well-studied cell types and model organisms~\citep{AcunerOzbacan2011,Davey2023,Huttlin2021}. While high-resolution, quantitative experiments enable reliable and precise detection of protein interactions, including weak, transient ones, their expense prohibits large-scale application~\citep{AcunerOzbacan2011}. Computational methods have the potential to inexpensively fill in existing gaps and yield a more comprehensive picture of protein interaction networks. Deep learning methods in particular can leverage both heterogeneous data and basic physical knowledge about protein interactions to model diverse molecular recognition mechanisms. This makes them well-suited for this task compared to other modeling techniques. On the one hand, molecular simulations rooted in statistical mechanics offer detailed physical insights into protein interaction specificities, thermodynamics, and kinetics, but are too computationally expensive to apply proteome-wide~\citep{Spiga2014,Wang2022_mdreview}. On the other hand, purely bioinformatic approaches predict interactions based on known sequence and structural homology or functional associations, but have difficulty generalizing to poorly characterized regions of interaction space~\citep{Tsuchiya2022,Peng2017,Xue2015,Kelly2008}.

In this review, we first describe how deep neural networks have been tailored to model protein interactions and, in particular, how the choice of inputs provided and their internal model representations, target outputs, and network architecture determine the types of interactions modeled and biological questions addressed. We then report on recent progress in using deep learning to accomplish three long-term goals: (1) discovery of protein interactions that have yet to be experimentally detected, (2) elucidation of the residues, molecular surfaces, and atomic structures that underpin protein interactions, and (3) design of protein binders and oligomeric assemblies for basic and translational applications. Throughout, our focus is on approaches suitable for bottom-up inference (and design) of interaction networks from physical and molecular determinants. Remarkable successes have been recently achieved using deep learning to structurally resolve protein interactomes~\citep{Burke2023,Humphreys2021,Gao2022,OReilly2023} and computationally design picomolar affinity binders~\citep{Vazquez2022}. Nonetheless, generalizing interaction prediction and design beyond sequences and structures well-represented in training data remains a challenge~\revision{\citep{Bernett2023,Dunham2022,Tsishyn2023}}. We conclude by discussing promising ideas across multiple aspects of model development: from compiling larger, higher-quality datasets to better incorporating biophysical priors within network architectures and integrating varied training schemes. We believe these ideas will help overcome existing challenges, including predicting interaction affinities and variant effects at scale, and expand the scope of current methods to include protein complexes mediated by highly transient interactions, intrinsically disordered regions, and post-translational modifications.

\section{Properties of protein interactions inform modeling approaches}

Geometric and chemical complementarity of interacting surfaces often plays a defining role in molecular recognition. These functional interfaces are more evolutionary conserved than other surface sites, and their complementarity is maintained through co-evolution of residues on binding partners~\citep{Ghadie2018}. Yet, within this basic framework, the specific biophysical mechanisms responsible for molecular recognition are highly diverse~\citep{AcunerOzbacan2011,Kastritis2013}. Such diversity makes it challenging to compile a set of comprehensive and predictive rules from first principles to describe dimeric protein interactions, let alone how combinations of pairwise interactions assemble proteins into higher-order complexes (Fig.~\ref{fig:ppi_types}B). In this case, it can be advantageous to use statistical approaches to capture the physicochemical, structural, and evolutionary patterns underpinning protein interactions directly from data. Neural networks accomplish this by learning to transform raw input representations of proteins (\textit{e.g.,} sequences, evolutionary histories, 3-dimensional (3D) structures and/or molecular surfaces) into abstract representations optimized for prediction or design of interactions. This avoids the need to identify and manually handcraft features of protein biophysics and evolution responsible for binding, an exercise based on still incomplete domain knowledge. Instead, well-established principles underlying molecular recognition can be incorporated into neural network primitives that transform raw protein inputs into learned representations. As a simple example, the arbitrary assignment of `receptor' and `ligand' labels to two globular proteins has no bearing on their ability to interact, and this insensitivity to label permutations can be incorporated into the network architecture by using order-invariant operations to process the input representations of all proteins ~\citep{Chen2019,Huang2023,Sledzieski2021}.

\begin{figure*}[htb!]
    \centering
    \includegraphics[width=438pt]{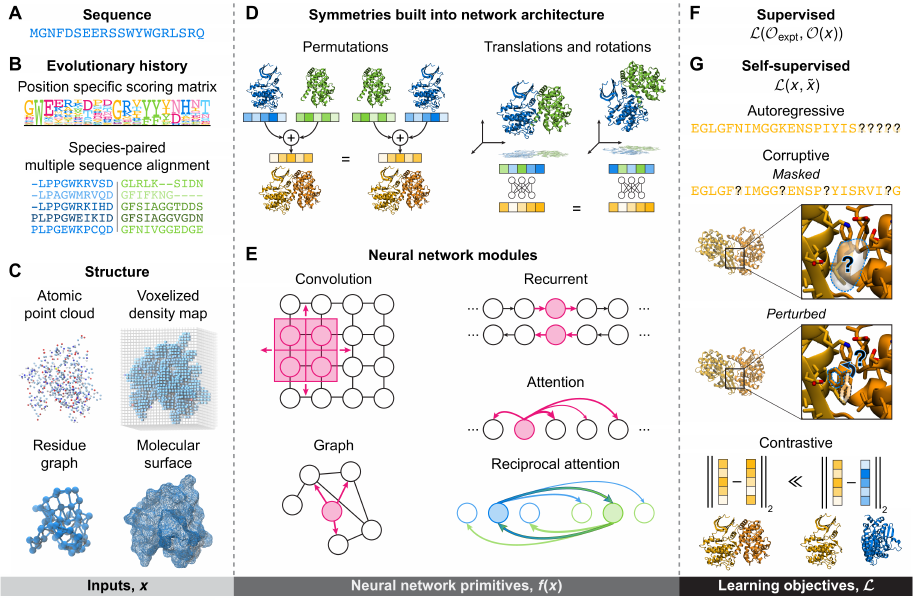}
    \caption{\textbf{Components of deep learning models of protein interactions.} Deep neural networks learn to transform (A-C) raw protein inputs, $x$, using (D-E) various network primitives into abstract representations optimized to achieve (F-G) particular learning objectives, $\mathcal{L}$. (A) As input to the model, proteins can be described by their primary amino acid sequences. (B) Evolutionary information can be supplied in the form of a position specific scoring matrix, which contains statistics of residue conservation at each sequence position, or, for example, a paired multiple sequence alignment, which contains statistics of residue conservation and correlation across all positions within and between species-paired proteins. (C) Structural information can be supplied in the form of a point cloud containing atomic coordinates, a voxelized density map tabulating the volumetric density of atoms on a discrete 3D grid, a residue graph encoding residue-level features in nodes and inter-residue quantities such as distances and angles in edges, or a molecular surface mapping atomic features onto a triangulated mesh that outlines the protein's solvent accessible surface. (D) Neural network primitives are chosen with appropriate symmetries built into the architecture. For example, combining representations of two proteins with permutation invariant operations, such as element-wise addition and multiplication or generalized outer products, ensures invariance to input ordering. Group equivariant networks update representations of protein structures to ensure equivariance to global translations and rotations. (E) Neural network primitives are also chosen with inductive priors reflective of the underlying biophysics. Convolutional networks process inputs with a regular grid structure by translating the same filters over all patches of the input and thus capture short-ranged, translationally-invariant multibody interactions. Recurrent networks process sequentially-ordered, variable-length inputs by iteratively applying the same transformations to each element and can be stacked to processes inputs in both directions. Graph neural networks process inputs that lack a regular grid-like structure or inherent ordering but have a fixed topology defined based on known relationships between elements, such as distances and angles between atoms in molecular structures. Representations of each node (and edge) are updated based on their connections to neighboring nodes. Attention-based networks explicitly consider all pairwise connections between elements and infer relationships between them during training, thus allowing for variable, flexible topologies and capturing of both short- and long-ranged interactions. In a self-attention primitive, representations of each element are updated by considering all other elements within the same input. In a reciprocal cross-attention primitive specifically formulated for protein--protein interactions, representations in one protein are updated based on those in the other protein according to attention values computed from both representations. To learn network parameters, models are trained through (F) supervised learning, in which the model is optimized to predict a target observable from the input, $\mathcal{O}(x)$, that matches its experimental value, $\mathcal{O}_{\rm expt}$. (G) Alternatively, models can be trained through self-supervised learning, in which a model is optimized to accomplish a proxy task constructed from the input data itself using an objective function defined over input data alone, $\mathcal{L}(x, \tilde{x})$. Among self-supervised approaches, autoregressive models learn to predict the next amino acid in a sequence given a truncated input, $\tilde{x}$. Models tasked to reconstruct protein sequences or structures from corrupted inputs, $\tilde{x}$, for instance with masked residues or perturbed side-chain orientations, learn contextual residue- or atomic-level representations. Contrastive learning optimizes the learned internal representations instead of a target output function. This is often accomplished by maximizing the similarity, such as measured by $\ell_2$ norm or cosine similarity, between representations of interacting protein pairs and minimizing that of non-interacting pairs.}
    \label{fig:dl_primitives}
\end{figure*}

All aspects of a neural network model, from the choice of inputs and biophysical priors to the training scheme used (Fig.~\ref{fig:dl_primitives}), influence how the underlying biological problem is posed and encoded within the model.
For example, protein--peptide interactions, unlike dimeric protein--protein interactions, naturally admit `receptor' and `ligand' labels, and this asymmetry motivates the use of two distinct network components for encoding protein and peptide inputs within a single model~\citep{Lei2021,Li2023,Abdin2022}. In a more complex setting, neural networks developed for paired protein inputs cannot model higher-order oligomers, and, when trained on complexes decomposed into direct and indirect pairwise interactions~\citep{Mosca2013_interactome3d,Peng2017}, will fail to capture contextual effects that emerge from multiway interactions. Conversely, many higher-order complexes, including ion channels, metabolic enzymes, and membrane remodeling proteins, adopt symmetric structures belonging to cyclic, dihedral, cubic, or helical point groups, and symmetry can be exploited to model and design these large complexes without explicitly representing all equivalent units, reducing computational cost~\citep{Li2022_UniFold_Symmetry,Watson2023}. Indeed, tailoring models to a particular type of protein interaction, even if it imposes practical constraints, is often valuable. In the following sections, we discuss the implications of some of the key choices encountered when building deep learning models of protein interactions.

\subsection{Deep learning models operate on protein sequences, structures, and evolutionary histories}

Proteins can be presented to models in multiple complementary ways (Fig.~\ref{fig:dl_primitives}A-C). Their individual amino acid sequences provide minimal descriptions amenable to modeling all types of interactions. Including evolutionary and structural information can facilitate predictions by exposing patterns of co-evolution among residues and potential surface complementarity, respectively. The former is most useful for interactions under strong selective pressures, including permanent, obligate interactions and protein hubs~\citep{Ghadie2018}, while the latter for interactions occurring with limited conformational changes. In contrast, short peptidic motifs are both structurally and evolutionary plastic; this makes them ideal for constructing and adapting complex regulatory programs that involve multiple binding partners~\citep{VanRoey2014} but potentially limits the utility of structural and evolutionary information in modeling them. Indeed, both the availability of a given data modality (or feasibility of acquiring it) and the prediction objective inform the decision to explicitly provide this data or to task the model with implicitly inferring it. For example, models can learn to transform unbound monomeric (apo) structures into a structure of the bound complex~\citep{Ganea2022,Ketata2023,Sverrisson2022}, but providing apo structures as inputs may bias predictions towards proteins that maintain their unbound conformations; alternatively, complex structures can be inferred directly from input protein sequences, implicitly modeling conformational changes upon binding~\citep{Evans2022}.

How particular information is provided to the model is a key question. For evolutionary information, residue conservation can be supplied as a position-specific scoring matrix (PSSM), which tabulates the empirical probabilities of every amino acid independently occurring at each residue position (Fig.~\ref{fig:dl_primitives}B). Multiple sequence alignments (MSAs) relax position independence and provide a comprehensive account of evolutionary variation, conservation, and correlation across all residue positions. While powerful, constructing high-quality MSAs and, especially, species-paired MSAs (Fig.~\ref{fig:dl_primitives}B) that capture inter-protein residue co-evolution (that may be indicative of a protein--protein interaction) can be challenging~\citep{Bryant2022_FoldDock}. In both PSSMs and MSAs, the evolutionary information encompassing a single protein family is presented explicitly to the model. Protein language models provide an alternate approach in which patterns observed across millions of (unrelated) protein sequences are distilled into a single model that attempts to implicitly capture all of known protein evolution~\citep{Alley2019,Rives2021,Lin2023,Chen2023_trimo}. Using a protein language model to embed a sequence also avoids incurring the computational costs of storing and searching large sequence databases to construct explicit MSAs.

For structural information, atomic point clouds~\citep{Krapp2023}, density maps~\citep{Townshend2019,Kozlovskii2021}, residue graphs~\citep{Abdin2022,Fout2017,Li2023}, or molecular surfaces~\citep{Gainza2020,Dai2021} can be provided as input to the model (Fig.~\ref{fig:dl_primitives}C). Point clouds describe the detailed arrangement of atoms in 3D space, but do not explicitly account for excluded volume, which leads to short-ranged repulsion between molecules. Density maps provide a voxelized representation in which electron density, for example, can be directly encoded, at the cost of expending memory and computations on unoccupied regions of space. Residue graphs, in which nodes correspond to the C$\alpha$ positions of residues, provide efficient coarse representations of backbone configurations but typically lack explicit descriptions of side chain positions, which may significantly reorient upon binding. Molecular surfaces represent proteins exclusively by exterior residues that may engage in direct physical contact with other molecules; however, as surface-based representations have limited capacity to capture internal residues and their impact on binding, they can struggle to capture collective conformational changes, such as coordinated by distributed allosteric networks.

\subsection{Deep learning models encode properties of protein interactions into abstract representations}

Neural network primitives with different inductive priors can be used, individually or in combination, to encode protein input data (Fig.~\ref{fig:dl_primitives}D-E). These primitives are typically chosen based on their suitability for different protein representations and ability to capture aspects of the underlying biophysics. Convolutional networks specialize in processing inputs with regular grid structures, including 1D sequences and voxelized 3D density maps, by reusing the same set of parameters to process grid patches. They can capture local, translationally-invariant patterns such as short peptide-binding motifs within proteins~\citep{Wardah2020,Kozlovskii2021,Lei2021}. By virtue of their densely connected filters, convolutional networks also excel at capturing collective many-body interactions between elements within a patch, but their reliance on fixed receptive fields limits the length-scale of the interactions that they can model in practice (larger receptive fields lead to rapid increases in parameter count and computational cost). By comparison, recurrent networks, which sequentially process elements in a 1D or higher-dimensional sequence, do not have explicit length-scale constraints and can in principle operate on arbitrarily long sequences. However, they have been empirically found to lose associations between far away elements, thus exhibiting an implicit locality bias. Furthermore, in contrast to most other neural network primitives which comprise numerous independent operations and are thus highly parallelizable, the sequentiality of recurrent networks makes them less suited to modern computing hardware.

Structured data lacking a regular grid-like topology or an inherent ordering, such as molecular structures~\citep{Fout2017} or protein interaction networks~\citep{Gao2023}, are more naturally modeled with graphs that explicitly encode known relationships. For instance, by encoding protein structures as graphs, neural networks can leverage existing biophysical knowledge to identify protein--protein interaction sites~\citep{Xie2022} and rigidly dock proteins~\citep{Sverrisson2022} based on their spatiochemical complementarity. However, such networks cannot explicitly represent protein--protein contacts learned during training or discovered at prediction time since they maintain the fixed topology of their inputs (they may however implicitly encode such information). Graph-based protein representations can include nodes for each heavy atom and model spatial relationships between atoms using binary or distance-weighted edges~\citep{Tubiana2022,Liu2021_GeoPPI}. Information then flows through the network according to the connectivity of the graph using permutation-invariant operations to update the representations of each node and edge (Fig.~\ref{fig:dl_primitives}E). As a result, like convolutional and recurrent networks, graph neural networks exhibit a spatial locality bias towards interactions between atoms or residues within the same physical neighborhood over ones distant in a protein's structure.

An advantage of graph-based representations, particularly for protein structure, is their invariance to global rotations and translations (Fig.~\ref{fig:dl_primitives}D); in the language of group theory, these are known as SE(3) transformations. SE(3)-invariance is a desirable property because how a protein is computationally oriented in 3D space has no bearing on its behavior. Graph neural networks achieve SE(3)-invariance by encoding geometrical aspects of structure using exclusively relative quantities (\textit{e.g.,} inter-residue lengths and angles) that do not vary with global position or orientation. By comparison, standard convolutional networks operating on voxelized densities are sensitive to SE(3) transformations unless trained using data augmented with arbitrary rotations to teach the model to recognize patterns in any orientation~\citep{Townshend2019,Kozlovskii2021}. However, graph neural networks are at a comparative disadvantage to convolutional networks in capturing many-body interactions. To harness the strengths of both, a parallel line of research has focused on directly endowing convolutional networks with awareness of SE(3) transformations. One approach is to parameterize the receptive fields of convolutional networks using SE(3)-equivariant mathematical functions, such as spherical harmonics, resulting in representations of atomic coordinates that deterministically and predictably change under coordinate transformations~\citep{Smidt2021}. Tailoring architectures to respect symmetries inherent to geometric structures, including graphs, point clouds, and surfaces, forms a subfield of deep learning, geometric deep learning~\citep{Bronstein2021}, that is increasingly being used to model protein interactions~\citep{Gainza2020,Ganea2022,Krapp2023,Dai2021,Huang2023,Si2023_PLMGraph-Inter}. By incorporating strong geometrical priors, SE(3)-equivariant networks~\citep{Geiger2022} more faithfully represent physical systems and molecular interactions; however, they can be technically challenging to implement efficiently.

As previously stated, graph neural networks prescribe a specific structure to the inputs, which can make them unsuitable when this structure is unknown or changing. To avoid this limitation, attention-based networks learn the relationships between input elements over the course of training (strictly-speaking, an attention-based network is analogous to a fully-connected graph in a graph neural network). Because all pairwise connections are considered within an attention primitive, information deemed valuable for solving a particular objective flows directly between input elements irrespective of their distance (Fig.~\ref{fig:dl_primitives}E). Although the required all-to-all computation is expensive, attention-based networks, unlike the primitives described earlier, have effectively unlimited receptive fields and can encode short- and long-ranged interactions equally well. These capabilities, combined with the minimal incorporation of inductive assumptions, allow the structure and inherent properties of an input to emerge during model inference, for example, to realize the dogma `sequence determines structure' and predict atomistic complex structures from protein sequences (although in practice, existing models require MSAs, not just individual sequences~\citep{Jumper2021,Evans2022}). Due to their expressivity, attention-based neural networks are increasingly used to capture complex and long-ranged interdependencies between residues within and across interacting proteins~\citep{Lei2021,Hosseini2022,Tsukiyama2022,Wang2022_protein_peptide_binding_residues}. Specific formulations of attention have also been developed to explicitly encode the inherently reciprocal relationships between two interacting proteins~\citep{Abdin2022,Baranwal2022} and to encode spatial relationships between residue pairs in a manner that emulates a defining property of distances in Euclidean geometry, namely the triangle inequality~\citep{Jumper2021,Wu2022_omegafold}.

\subsection{Deep learning models acquire meaningful protein representations through supervised and self-supervised learning}

Neural networks are typically optimized end-to-end, meaning that all model parameters are tuned simultaneously to maximize the final objective function of the model (output end). This induces the model to learn to encode raw data (input end) into abstract representations tailored for its particular objective function. In supervised learning, objective functions are defined over quantities dependent on, but distinct from, model inputs (Fig.~\ref{fig:dl_primitives}F); for example, an objective could be defined to minimize the error between experimentally measured affinities of protein--protein interactions and ones predicted given input protein sequences. However, experimental determination of target observables for all training examples can be costly and time-consuming, whereas input data modalities are often more readily available. Self-supervised learning is an alternative approach in which input data alone is used to create proxy tasks and associated objective functions for the model to optimize (Fig.~\ref{fig:dl_primitives}G). With suitably chosen tasks, this process can lead the model to learn abstract representations that are generically useful (\textit{i.e.}, that encode intrinsic properties of the data that facilitate reasoning for a broad array of tasks). For protein interaction modeling, large, existing databases of protein sequences or structures can be leveraged in this way to learn useful representations even if their interaction partners are unknown.

One self-supervised objective tasks models with predicting the next amino acid in a protein sequence (Fig.~\ref{fig:dl_primitives}G). The resulting models acquire the ability to generate novel protein sequences, implicitly learning principles that underlie natural protein stability and function~\citep{Ferruz2022}. Such autoregressive models are valuable for \textit{de novo} sequence design but assume that the identity of each residue depends exclusively on ones upstream of it and, thus, do not capture a residue's full sequence context. An objective devised to avoid this limitation instead tasks models with recovering the amino acid identities of masked or corrupted residues based on the rest of the protein sequence or structure~\citep{Lin2023,Liu2021_GeoPPI,MohseniBehbahani2023,Mahajan2023}. Through such a masked language modeling objective, protein language models acquire contextual representations of protein sequences that capture evolutionary, physicochemical, and structural information~\citep{Rives2021} applicable for solving downstream tasks ranging from protein interaction~\citep{Hallee2023,Sledzieski2021,Chen2023_trimo} and variant effect prediction~\citep{Unsal2022} to complex structure prediction~\citep{Lin2023} and binder design~\citep{Hie2022,Bhat2023}. To learn contextual representations of protein structures that capture atomistic details pertinent to molecular recognition mechanisms, one data corruption approach leverages the insight that side chain rearrangements at protein complex interfaces can disrupt key stabilizing interactions by tasking models with reconstructing perturbed side chain orientations. The resulting learned representations encode geometric and physicochemical information predictive of mutational effects on binding affinity~\citep{Liu2021_GeoPPI}.

While the above approaches encourage models to encode fundamental properties of protein sequences or structures, they do not explicitly require a sophisticated understanding of inter-protein relationships that determine, for example, homology and interaction specificity. To capture characteristic commonalities and differences between data points, a third type of self-supervised objective tasks models with maximizing the similarity of representations for related inputs over unrelated ones (Fig.~\ref{fig:dl_primitives}G). Such contrastive learning objectives have been used to encourage separation between the abstract representations of interacting versus non-interacting protein pairs and facilitate protein interaction prediction~\citep{Wang2022_protein_peptide_binding_residues,Palepu2022,Gainza2020}. Furthermore, by combining objectives that require distinct but complementary methods of reasoning over inputs, models can acquire more holistic representations of proteins that are generally useful for solving complex, varied tasks~\citep{Chen2023_trimo,Zhang2023}. For this reason, self-supervised objectives have also been integrated within supervised training schemes to fully leverage all types of data available for modeling protein interactions and complex structures~\citep{Jumper2021,Evans2022,MohseniBehbahani2023,Wang2022_protein_peptide_binding_residues}.

\section{Efforts to discover, elucidate, and design protein interactions using deep learning}

\begin{figure*}[htb!]
    \centering
    \includegraphics[width=438pt]{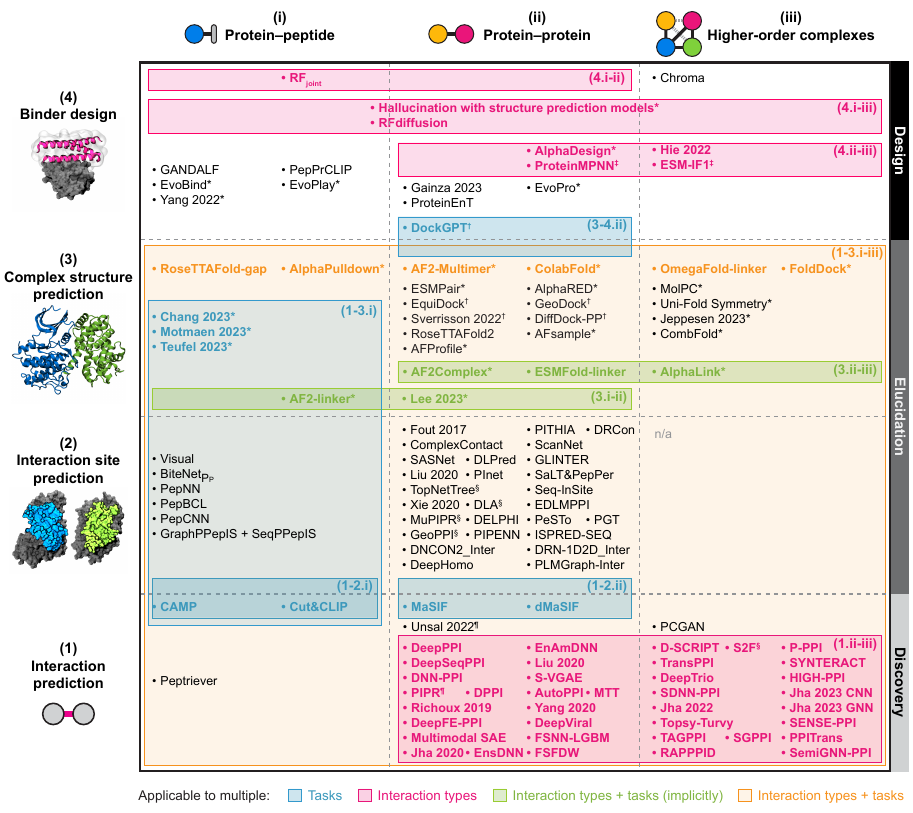}
    \caption{\textbf{Ecosystem of deep learning methods for modeling protein interactions.} Deep learning models have been developed to predict (1) interactions, (2) interaction sites, and (3) atomistic complex structures and to (4) design both direct physical interactions for (i) protein--peptide and (ii) protein--protein dimers as well as (iii) high-order complexes, including their complete pairwise decompositions into direct physical interactions (solid black lines) and indirect interactions (dashed gray lines). Boxed and colored models are applicable to multiple tasks and/or types of protein interactions and labeled accordingly (\textit{e.g.,} models labeled (1-2.i) predict interactions and interaction sites for protein--peptide interactions). All models for complex structure prediction implicitly predict interaction sites as well. References for all models are provided in the appendix. \textdagger\ Models for protein docking. \textdaggerdbl\ Models for designing protein sequences given a complex structure. $\mathparagraph$ Models for predicting binding affinity. $\mathsection$ Models for predicting changes in binding affinity upon mutation. * Models based on AlphaFold2.}
    \label{fig:task_vs_type}
\end{figure*}

Deep learning models have been developed to predict the probability of experimentally detecting an interaction between two proteins, the residues and molecular surfaces that mediate an interaction, and the atomistic structure of a bound complex. These models form the foundation of a growing ecosystem of deep learning methods for discovering novel protein interactions, elucidating their physical mechanisms, and designing binders to precisely modulate their function. In the following sections, we describe the main types of approaches that have been employed to address each of these biological questions, while a comprehensive list of models is provided in Fig.~\ref{fig:task_vs_type}.

\subsection{Discovering interactions}

\begin{figure}[htb!]
    \centering
    \includegraphics[width=213pt]{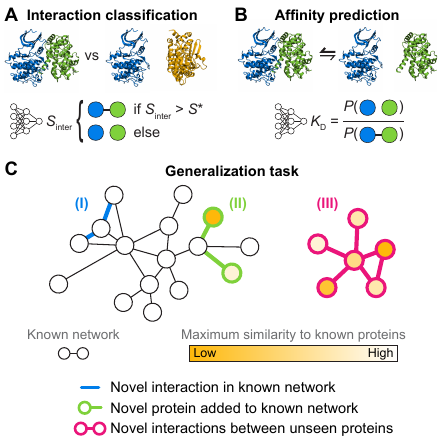}
    \caption{\textbf{Efforts to discover protein interactions employ deep learning methods to accomplish multiple tasks.} (A) Most methods classify a pair of proteins as interacting or non-interacting based on a predicted interaction score ($S_{\rm inter}$). If $S_{\rm inter}$ is greater than a chosen threshold $S^*$, the pair is labeled as interacting, otherwise it is labeled non-interacting. (B) Other methods quantify the affinity of an interaction between two proteins, specifically the dissociation constant ($K_{\rm D}$), which quantifies the relative probability of observing two proteins in their unbound monomeric state versus bound state. (C) Models are evaluated for their ability to generalize to proteins dissimilar from ones in their training sets (based on sequence or structural homology) in order to (I) impute new interactions among a set of known interacting proteins; (II) identify interactions between a protein unseen in the training set and known PPI network; and (III) extrapolate to interactions between pairs of unseen proteins.}
    \label{fig:discovery_subtasks}
\end{figure}

Scalable computational models suitable for screening large sets of possible protein interactions have been developed to identify novel interacting protein pairs. Methods for this task have typically been devised to complement high-throughput experimental efforts to fully map protein interactomes~\citep{Luck2020,Huttlin2021}. Analogous to the qualitative readouts obtained from these experiments, which indicate only the detection of an interaction and not its affinity, these models almost universally formulate interaction prediction as a binary classification problem: two proteins either interact or do not based on a predicted interaction score (Fig.~\ref{fig:discovery_subtasks}A).
In the case of purely bioinformatic methods, predicted interaction scores derive from homology (based on sequence or structural similarity) to an experimentally-validated interacting pair~\citep{Petrey2023,Zhang2012_PrePPI_original,Peng2017,Kelly2008}. As such, these methods depend on the availability of homologous templates to make reliable inferences. The increase in high-confidence structural models predicted by deep learning models, whose application to complex structure prediction is discussed in a latter section, has enabled homology-based approaches to expand their applicability while maintaining their capacity for rapid screening~\citep{Petrey2023}.

Deep learning interaction predictors avoid the need for input homologous templates altogether, and instead make inferences based on raw input representations of individual proteins. They may, nevertheless, learn to implicitly encode homology to known protein interactors used for training~\citep{Bernett2023,Yang2023}. Training models for binary interaction prediction requires experimentally-validated positive (interacting) and negative (non-interacting) examples. While high-throughput experimental methods have detected over 100,000 positive interactions to-date~\citep{Luck2020,Huttlin2021}, analogous methods to confirm negatives at scale are lacking. As a result, there is a relative paucity of experimentally-validated negative examples, of which we estimate there are roughly 6,000 curated instances within databases~\citep{Blohm2014,Trabuco2012}. This is in stark contrast to estimates that, among human and yeast proteomes, negatives exceed positive interactions by 1,000 fold~\citep{Trabuco2012,Luck2020}. One standard approach to circumvent the scarcity of experimentally-validated negatives and better match the expected sparsity of interaction networks is to train models with synthetically generated negative examples: Assuming that all protein pairs not observed in experimental datasets are negatives, they can be uniformly sampled to generate a much larger number of negatives than positives~\revision{\citep{Hamp2015_evolutionary_profiles_improve,Hashemifar2018,Sledzieski2021,Szymborski2022,Volzhenin2023,Lei2021,Dunham2022}}. Another approach to generate synthetic negatives assumes that proteins localized to different cellular compartments are unlikely to interact~\citep{Ben-Hur2006,Du2017,Zhang2019_multimodal_deep_representation_learning,Yang2020_SVGAE,Hallee2023}. Other approaches do exist with different inductive priors~\citep{Hallee2023,Ben-Hur2006}, and this remains an active area of research.

Under the umbrella of `interaction prediction' exist three specific tasks for which models have been developed (Fig.~\ref{fig:discovery_subtasks}C): (I) imputation of novel interactions within a known, experimentally-validated protein interaction network, (II) addition of new proteins, which currently lack experimentally-validated interactions, to a known network, and (III) construction of novel interaction networks composed of new proteins. A model's suitability for each of these tasks is determined through assessment on a test set of experimentally-validated interactions constructed to differ from the training set in a manner that mimics the desired use case: (I) interactions (edges in the network) are split such that individual proteins may occur in both training and test sets, (II) individual proteins are split into training and test sets such that all test interactions involve at least one test protein absent from the training set, and (III) individual proteins are split such that all test interactions involve only test proteins~\revision{\citep{Park2012,Hamp2015_more_challenges,Dunham2022,Szymborski2022}}. Models consistently excel at imputing interactions (task I), achieving accuracies above 95\%~\citep{Bernett2023}. \revision{As demonstrated by S-VGAE~\citep{Yang2020_SVGAE}, TAGPPI~\citep{Song2022}, SGPPI~\citep{Huang2023}, and the models developed by~\citeauthor{Jha2023_PPI-Graph-BERT}~\citep{Jha2020,Jha2023_analyzing_effect_of_multi_modality,Jha2023_PPI-Graph-BERT},} including structural information as input in addition to raw sequences can slightly improve accuracies by roughly $1-4$\%. However, sequence-based models using representations obtained through self-supervised pre-training~\citep{Chen2019,Yao2019,Dong2021,Sledzieski2021,Hallee2023,Palepu2022,Anteghini2023}\revision{, specifically Peptriever~\citep{Gurvich2023} and PPITrans~\citep{Yang2023} models,} match or outperform structure-based models in recent evaluations. This may be because representations obtained through self-supervised learning implicitly capture structural information (in addition to evolutionary information). Indeed, large protein language models have been shown capable of directly encoding atomic-level structural information, such as intra-protein residue contacts~\citep{Lin2023,Rao2021}.

Predicting interactions involving novel protein(s) (tasks II and III) requires models to generalize to proteins outside their training sets. Such capabilities are especially key for constructing interaction networks that include understudied or low-abundance proteins and those of non-model organisms~\citep{Volzhenin2023}. When \revision{RAPPPID~\citep{Szymborski2022}, 
 CAMP~\citep{Lei2021}, PPITrans~\citep{Yang2023}, and SENSE-PPI~\citep{Volzhenin2023} models were} evaluated on these more challenging tasks, classifier effectiveness, as measured by the areas under the receiver operating characteristics curve (AUROC) and precision-recall curve (AUPR), dropped by roughly 10\% and 20\% on interactions involving one (task II) or both proteins (task III) being absent from the training set, respectively, with models recalling around 70\% of known interactions~\citep{Yang2023}. Furthermore, because protein interaction surfaces are evolutionarily conserved~\citep{Ghadie2018}, proteins with unique but related sequences may nevertheless engage similar binding partners, which can be readily predicted by homology alone. To assess how well models generalize to evolutionary distant proteins, within both these evaluation schemes, models have been assessed on test proteins with varying levels of homology to proteins in the training set based on sequence identity~\citep{Szymborski2022,Lei2021} or the species in which interactions occur~\citep{Yang2023,Volzhenin2023,Huang2023,Hashemifar2018}. \revision{RAPPPID~\citep{Szymborski2022}, CAMP~\citep{Lei2021}, PPITrans~\citep{Yang2023}, and SENSE-PPI~\citep{Volzhenin2023} models} perform nearly equally well on interactions involving proteins that range in sequence identity from 100\% to 40\% to a training example and, concordantly, test species with median sequence identities greater than 40\% to a training species, but performance drops drastically on protein pairs less than 30\% sequence identical to all proteins in the training set. The most successful methods for these two tasks employ representations from large protein language models\revision{, as done in PPITrans~\citep{Yan2023} and SENSE-PPI~\citep{Volzhenin2023} models}, enhanced regularization techniques\revision{, as done in RAPPPID~\citep{Szymborski2022}}, or multitask learning strategies\revision{, as done in CAMP~\citep{Lei2021},} to reduce overfitting on the training set and improve model generalizability; if such techniques are not used, models have been shown to randomly classify protein pairs absent from training sets as interacting or non-interacting~\revision{\citep{Szymborski2022,Yang2023,Volzhenin2023,Bennett2023,Dunham2022}}.

While extrapolating outside the training set is crucial for discovering novel interactions, the capability to quantitatively predict interaction affinities (Fig.~\ref{fig:discovery_subtasks}B) with high sensitivity to single residue mutations is key to unraveling variant effects. Scores predicted by binary classifiers \revision{DPPI~\citep{Hashemifar2018}, CAMP~\citep{Lei2021}, and that developed by~\citeauthor{Motmaen2023}} show modest correlation with affinity (Pearson $R \approx 0.7$) but vary from experimental values by roughly a factor of three. Such error is greater than that typically reported for low-throughput experimental measurements, which commonly differ by only a factor of two across technical replicates, methods, or labs~\citep{Jankauskaite2019}. Training quantitative regression models has been challenged by the limited number of curated affinity measurements compared to qualitative measurements: Only about $7,000$ affinities are available in the SKEMPI database for 345 complexes and their point mutants~\citep{Jankauskaite2019}. To date, only two deep learning methods\revision{, PIPR~\citep{Chen2019} and those developed by~\citeauthor{Unsal2022},} have been trained to quantify interaction affinities. By leveraging protein language models, which can directly encode variant effects on individual protein's functions~\citep{Rives2021,Meier2021}, \revision{the regression models of~\citeauthor{Unsal2022}} exhibit substantially increased correlation with experimental affinities (Pearson $R = 0.9$) than binary classifiers but their errors still exceed experimental ones.

Across all problem formulations, models predict different types of interactions (Fig.~\ref{fig:ppi_types}), and which types are predicted by a given model are largely determined by the choice of training data. In the case of protein--peptide interactions, \revision{the three models developed to date, CAMP~\citep{Lei2021}, Peptriever~\citep{Gurvich2023}, and Cut\&CLIP~\citep{Palepu2022},} have been trained specifically on direct physical interactions, almost exclusively derived from experimental structures in the Protein Data Bank (PDB) and curated into the PepBDB~\citep{Wen2019} and Propedia~\citep{Martins2021} databases. In the more general case of protein--protein interactions, training datasets have been derived from a variety of assays, ranging from \textit{in vitro} biophysical assays to \textit{in vivo} assays and ones based on cell extracts such as affinity purification mass spectrometry, with each providing evidence for a different interaction type(s). Many high-throughput assays, in fact, do not differentiate between direct physical interactions (Fig.~\ref{fig:ppi_types}A) and indirect interactions mediated by intermediary proteins (Fig.~\ref{fig:ppi_types}B)~\citep{AcunerOzbacan2011}, nor do they distinguish between true dimers, whose stability does not require any additional binders, and pairs of proteins that are in direct physical contact only as part of a larger complex. Models trained on amalgamations of these experimental sources will invariably conflate different types of interactions. In instances where ambiguity is undesirable, datasets can be limited to experimental methods that provide unequivocal evidence for the sought interaction type, a curation step already performed by some databases. For example, the IntAct database~\citep{delToro2022} provides a filter for true dimers and has been used to create training and evaluation datasets for the MTT model~\citep{Dong2021}; the HINT database~\citep{Das2012} provides a filter for direct physical interactions, including true dimers and more generally direct interactions derived from yeast two-hybrid assays\revision{; and the BioGRID database~\citep{Oughtred2021} has been filtered for direct physical interactions by~\citeauthor{Dunham2022} to create datasets for benchmark evaluations.}

Faithfully predicting higher-order complexes, on the other hand, remains challenged both by standard data curation procedures, which record known interactions in a pairwise manner, and by constraints imposed by current model architectures, which exclusively process pairs of proteins as input. These practices are suitable for dimers but require pairwise decomposition of complexes with three or more proteins. Such a decomposition may be physically appropriate for higher-order complexes that assemble from independent dimers but would be inaccurate for obligate trimers, for example. Notable exceptions to these practices, include, on the data side, the cataloguing of over 5,000 complexes and their subunit stoichiometries within the CORUM~\citep{Tsitsiridis2022} and Complex Portal~\citep{Meldal2022} databases. On the model side, deep learning methods for complex structure prediction, described in a latter section, can accommodate a variable number of input monomer sequences~\citep{Evans2022,Baek2021,Mirdita2022,Bryant2022_FoldDock,Bryant2022_MolPC,Gao2022}. However, they require \textit{a priori} knowledge of complex stoichiometry, and this, in combination with the increased computational costs and memory requirements associated with modeling the structure of an increasing number of subunits, makes existing methods impracticable for screening putative higher-order complexes at scale. One deep learning method\revision{, PCGAN~\citep{Pan2023}}, takes an entirely different approach and predicts which protein nodes in an existing interaction network assemble into a higher-order complex; such an approach could be applied to interaction networks predicted using binary interaction classifiers in order to discover novel higher-order complexes.

\subsection{Elucidating interaction mechanisms}

\subsubsection{Predicting interaction sites}

\begin{figure}[htb!]
    \centering
    \includegraphics[width=213pt]{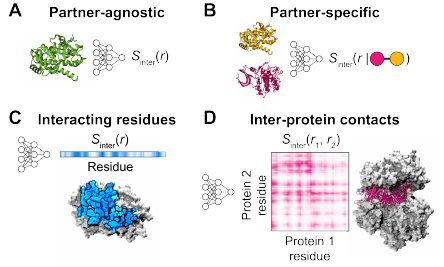}
    \caption{\textbf{Efforts to elucidate the molecular surfaces that mediate protein interactions employ deep learning methods to accomplish multiple tasks.} Most methods for interaction site prediction classify residues ($r$) of proteins as interacting or non-interacting based on a predicted interaction score ($S_{\rm inter}$). (A) Partner-agnostic models predict scores that are independent of a specific binding partner, whereas (B) partner-specific models predict scores conditioned on a particular interaction. (C) Interacting residues on an individual protein are predicted by all partner-agnostic models and partner-specific protein--peptide models, whereas (D) inter-protein residue contacts between a specific pair of residues are predicted by most partner-specific protein--protein models.}
    \label{fig:interact_site_subtasks}
\end{figure}

Identifying protein patches that physically contact other proteins is both practically useful for rationally designing interaction modulators and helps improve our understanding of the physicochemical and structural principles that underpin interaction specificity and affinity. Methods for this task fall into two broad categories: partner-agnostic ones that predict protein sites engaging one of possibly many partners (including drugs) (Fig.~\ref{fig:interact_site_subtasks}A), and partner-specific methods that predict sites engaging a particular partner (Fig.~\ref{fig:interact_site_subtasks}B).
The definition of what constitutes an interaction site can vary but is generally based on inter-protein atomic distances~\citep{Xue2015}, derived from experimental protein complex structures. Labels are typically assigned on a per-residue basis, even though a set of residues within a contiguous surface patch collectively determines interaction affinity and specificity~\citep{Kastritis2013,AcunerOzbacan2011}. As one way to account for such correlations between residues within a patch, \revision{the residue-level predictor ScanNet~\citep{Tubiana2022}} spatially smooths predicted interaction propensities. Since interacting residues provide evidence of a direct physical interaction, multitask learning approaches have also been deployed for these two synergistic tasks\revision{, such as done in CAMP}~\citep{Lei2021}.

Most interaction site predictors are not developed to delineate how different residues or regions contribute to binding affinity and, thus, neither directly distinguish hot spot residues from others nor assess mutational effects on binding affinity. Indeed, when evaluated for \textit{in silico} alanine scanning, a standard approach to identify hot spot residues, per-residue scores derived from \revision{the binary classifier ScanNet~\citep{Tubiana2022}} show minimal correlation with experimental changes in affinity (Pearson $R = 0.2$). Models that do predict the effects of point mutations on binding affinity\revision{, GeoPPI~\citep{Liu2021_GeoPPI}, DLA~\citep{MohseniBehbahani2023}, TopNetTree~\citep{Wang2020_TopNetTree}, and MuPIPR~\citep{Zhao2020},} often leverage residue-level representations of complex structures extracted from self-supervised models\revision{. By performing self-supervised training on $\sim$100 times more complex structures than available matching affinity measurements, GeoPPI~\citep{Liu2021_GeoPPI} and DLA~\citep{MohseniBehbahani2023}} achieve root-mean-squared-errors less than the threshold typically used to define hot spots (binding free energy changes greater than 2 kcal/mol), but their correlation with experimental values remains modest (Pearson $R \approx 0.6$) and drops substantially on complexes with reduced sequence or structural similarity to those in their training sets (Pearson $R \approx 0.4$). Developing generalizable models to quantify variant effects at scale remains a challenge~\citep{Tsishyn2023}, exacerbated by the limited amount of experimental data available for both training and evaluating models.

Previous studies have found that partner-specific methods are generally more accurate than partner-agnostic ones, especially for identifying sites that mediate transient interactions~\citep{Xue2015,Abdin2022,Dai2021}. Among partner-specific methods, protein--peptide models tend to predict interacting residues on either the protein\revision{, such as done by PepNN~\citep{Abdin2022}, PepBCL~\citep{Wang2022_protein_peptide_binding_residues}, and PepCNN~\citep{Chandra2023},} or the peptide\revision{, such as done by CAMP~\citep{Lei2021}}, but not both (Fig.~\ref{fig:interact_site_subtasks}C). On the other hand, protein--protein models not only identify the residues that comprise interacting surfaces on both proteins~\citep{Dai2021}, but also typically predict the specific residue pairs that form inter-protein contacts (Fig.~\ref{fig:interact_site_subtasks}D), which altogether characterize how the proteins are oriented relative to each other within the complex~\citep{Townshend2019,Fout2017,Zeng2018,Xie2020,Sledzieski2021,Xie2022,Hosseini2023,Si2023_DRN-1D2D_Inter,Si2023_PLMGraph-Inter}. This simultaneous prediction of residues on both interaction surfaces, \revision{when it is done}, appears to improve prediction accuracy for both protein--protein~\citep{Liu2020,Xue2015} and protein--peptide interaction sites, \revision{as is the case for GraphPPepIS and SeqPPepIS~\citep{Li2023}}. The improvement may be due to collective effects only being evident when the entirety of the interfacial region(s) is considered.

Within both categories of methods, models using structural information generally outperform those using only protein sequences~\citep{Li2023,Abdin2022}. While not a concern for sequence-based models, the suitability of structure-based models for identifying novel interaction sites rests on their success in using unbound monomer structures as input; knowledge of a complex structure abrogates the need to predict interaction sites altogether. Complexes with experimental structures for which the corresponding apo structures are unavailable outnumber those for which they are available by over one hundred to one~\citep{Townshend2019}. As a result, many models have been trained on bound monomer structures extracted from complex structures, either alone or in combination with apo structures~\citep{Townshend2019,Gainza2020,Dai2021,Li2023,Abdin2022,Kozlovskii2021,Tubiana2022,Krapp2023,Si2023_PLMGraph-Inter}. These models can reliably describe interactions between proteins that maintain `near-rigid' structures upon binding: For proteins that undergo minimal conformational changes, amounting to only a few angstrom root-mean-square-deviation between bound and unbound structures, \revision{ScanNet~\citep{Tubiana2022} and PeSTo~\citep{Krapp2023}} identify interacting residues on apo structures with only minimally reduced accuracy compared to bound ones. However, these models typically report reduced performance on unbound structures, overall, and those of proteins involved in transient interactions, in particular, with classifier effectiveness, as measured by AUROC, dropping by $5-10\%$ \revision{for MaSIF~\citep{Gainza2020}, PInet~\citep{Dai2021}, and PeSTo~\citep{Krapp2023}}. Thus, they remain less suited for modeling interactions in which monomers undergo significant conformational changes upon binding~\citep{Gainza2020,Dai2021}. Such interactions may be better described by models that instead predict complex structures directly from sequence, avoiding the need for monomer structures altogether and implicitly accounting for conformational changes.

\subsubsection{Predicting complex structures}

\begin{figure}[htb!]
    \centering
    \includegraphics[width=213pt]{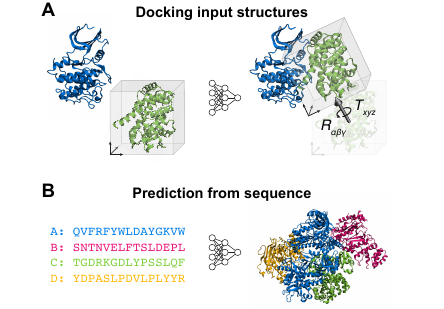}
    \caption{\textbf{Efforts to elucidate the atomistic structure of a protein complex using deep learning employ different strategies.} (A) Atomistic complex structures can be predicted by docking input monomer structures. Most commonly, rigid docking approaches are employed to translate and rotate one interactor into its bound position. Flexible docking approaches allow for additional conformational changes in each monomer. (B) Alternatively, atomistic complex structures can be predicted from the sequences of each constituent monomer and their stoichiometries in the assembly.}
    \label{fig:complex_struct_subtasks}
\end{figure}

Further mechanistic information about a protein interaction can be gleaned from its atomistic complex structure. When monomeric structures of each binding partner are available, docking algorithms can produce complex structures, conventionally through a search phase to sample different bound configurations (in addition to varied monomer conformations in flexible docking approaches) followed by scoring of each complex to select the optimal one (Fig.~\ref{fig:complex_struct_subtasks}A)~\citep{Li2022,Tsuchiya2022,Wodak2023}.
When interaction site information is unknown, the requisite global search and its computational expense often hinders successful docking. Focusing the search using deep learning methods for interaction site prediction, even when they are not perfectly accurate, can increase the success rate of conventional docking methods~\citep{Li2023,Si2023_DRN-1D2D_Inter,Si2023_PLMGraph-Inter}. End-to-end deep learning methods that avoid the search phase altogether\revision{, namely EquiDock~\citep{Ganea2022}, DiffDock-PP~\citep{Ketata2023}, GeoDock~\citep{Chu2023}, DockGPT~\citep{McPartlon2023}, and that developed by~\citeauthor{Sverrisson2022}}, offer more significant speedups over conventional docking methods. Ultimately, the success of docking methods depends on accurately scoring sampled complexes to identify the native pose among potentially many incorrect ones~\citep{Lensink2021}, and deep learning methods have been developed for this task as well~\citep{Renaud2021,MohseniBehbahani2022,Wang2021,Wang2020,Eismann2021,Han2023,Chen2023,Cao2020,Johansson-Akhe2021}.

Deep learning methods that predict atomistic complex structures from monomer sequences and their stoichiometries (Fig.~\ref{fig:complex_struct_subtasks}B) circumvent the need for input monomer structures. The unprecedented accuracy of AlphaFold2 (AF2) at predicting structures of individual protein chains~\citep{Jumper2021} prompted the development of approaches that use AF2 to predict complex structures. AF2 predicts atomistic structures by processing a protein's amino acid sequence, evolutionary history (in the form of an MSA), and, if available, structural templates of homologous proteins through a series of attention-based neural networks. Provided with monomer inputs connected either with a residue insertion gap~\citep{Baek2021,Bryant2022_FoldDock,Mirdita2022} or flexible glycine linker~\citep{Ko2021,Tsaban2022}, AF2 often successfully predicts complex structures, and an AF2 model developed specifically for complex structure prediction (AF2-Multimer)~\citep{Evans2022} exhibits superior accuracy~\citep{Evans2022,Bryant2022_FoldDock,Gao2022}: Out of 2,609 unique interfaces derived from over 4,000 complexes, 70\% of heteromeric and 72\% of homomeric interfaces are correctly predicted (as defined by DockQ score $\ge 0.23$~\citep{Basu2016}) by AF2-Multimer compared to only 43\% and 64\% by AF2 supplied with monomers connected with a flexible linker~\citep{Evans2022}. AF2-based approaches outperform conventional protein--protein and protein--peptide docking methods and exhibit almost three-fold higher success rates~\citep{Ghani2022,Yin2022,Akdel2022,Bryant2022_FoldDock,Johansson-Akhe2022,Shanker2023}, although further gains are possible with consensus approaches that combine AF2 with traditional, physics-inspired docking methods~\citep{Shanker2023,Johansson-Akhe2022,Ghani2022}).

Additionally, AF2 provides confidence metrics that can distinguish more accurately modeled complex structures from less accurate ones~\citep{Yin2022,Bryant2022_FoldDock,Gao2022,Johansson-Akhe2022,Zhu2023}, where accuracy is typically defined based on DockQ scores or metrics defined by the Critical Assessment of Predicted Interactions (CAPRI)~\citep{Lensink2021}. Moreover, AF2 confidence metrics can be used to distinguish directly interacting from non-interacting protein--protein~\citep{Gao2022} and protein--peptide pairs~\citep{Johansson-Akhe2022,Bret2023,Gurvich2023,Bryant2022_EvoBind,Teufel2023,Lee2023}. Evaluated on a classification benchmark of 672 protein--peptide interactions derived from experimental structures in the PDB, AF2-Multimer recalls 26\% at a false positive rate of 1\%~\citep{Johansson-Akhe2022} and, in fact, is matched only by one model developed specifically for this task---Peptriever~\citep{Gurvich2023}, which leverages self-supervised pre-training to jointly learn representations of proteins and peptides.

Despite the wide margin in accuracy between AF2-Multimer and state-of-the-art docking methods, AF2-Multimer predicted acceptable structures, defined based on metrics quantifying both overall fold and interface accuracy, for only 22 out of 41 complexes (54\%) in the blind, community-wide assessment conducted during the most recent Critical Assessment of protein Structure Prediction (CASP15)~\citep{Ozden2023}. Complexes with weak evolutionary signals, lacking structural templates, or assembled from many heterogeneous subunits and containing over 1,800 residues proved particularly challenging~\citep{Ozden2023,Zhu2023,Akdel2022}. These shortcomings were in part alleviated by optimizing MSA construction~\citep{Ozden2023}\revision{, such as done in AFProfile~\citep{Bryant2023_AFProfile} and ESMPair~\citep{Chen2022_ESMPair}}; increasing the number and diversity of predicted complex structures, for example, by increasing the number of iterations through the AF2-Multimer network (recycles) or randomly disabling neurons (dropout)~\citep{Johansson-Akhe2022,Wallner2023}; and assembling higher-order complexes from smaller interacting subcomponents individually predicted by AF2-Multimer\revision{, such as done in MolPC~\citep{Bryant2022_MolPC}, CombFold~\citep{Shor2023}, and the method developed by~\citeauthor{Jeppesen2023}}. However, none of these strategies individually yielded acceptable structures for all test complexes. The top ranking approach, which leveraged methods for constructing diverse and sensitive MSAs, successfully predicted 30 complexes (73\%) and was complemented by an approach employing increased sampling~\citep{Ozden2023}. Integrating available experimental information, specifically distance restraints obtained from methods such as cross-linking mass spectrometry, either to select and validate an optimal complex structure among those predicted~\citep{OReilly2023,Shor2023,Burke2023} or to directly bias the network's predictions\revision{, as done in AlphaLink~\citep{Stahl2023_AFLink-multimer},} promises further improvements.

Indeed, different types of protein interactions pose their own modeling challenges and motivate a variety of approaches to tackle them. Characteristic differences in the physical and evolutionary properties of protein--protein versus protein--peptide interactions warrant the use of different inputs or variations of AF2. For protein--protein interactions, combining paired MSAs (Fig.~\ref{fig:dl_primitives}B) with unpaired ones improves predictions with AF2-Multimer~\citep{Yin2022,Bryant2022_FoldDock}, but is often unnecessary to predict accurate complex structures when templates are used~\citep{Yin2022,Gao2022}. In contrast, MSAs for peptide interactors, being difficult to construct due to their short lengths, are typically not used~\citep{Ko2021,Tsaban2022,Johansson-Akhe2022,Chang2023,Shanker2023}, and using MSAs derived from their cognate proteins can even be detrimental~\citep{Bret2023}. In fact, including the full protein context surrounding peptidic binding motifs, which often reside within intrinsically disordered regions, substantially decreases prediction sensitivity and accuracy~\citep{Lee2023,Bret2023}. To address this limitation, a fragmentation approach, which decomposes full length proteins into constituent globular domains and peptidic motifs and removes disordered linker regions, has been devised to predict multidentate protein--protein interactions mediated by modular peptide-binding domains~\citep{Lee2023}. Interactions with limited evolutionary history, including rapidly evolving peptidic sites and antibody--antigen interactions, may be more suitably modeled with deep learning methods that predict structure directly from single sequences and avoid MSAs entirely; however, attempts using such methods\revision{, specifically ESMFold~\citep{Lin2023} and OmegaFold~\citep{Wu2022_omegafold},} with monomer inputs connected with a residue insertion gap or flexible linker have so far proven unsatisfactory compared to AF2-based approaches~\citep{Tsuchiya2022,Shanker2023,Zhu2023,Lin2023}.

By combining high-throughput experimental measurements of protein interactions with AF2 predictions, structurally resolved protein interaction networks can be constructed at proteome-scale. To-date, initial networks have been constructed for \textit{Saccharomyces cerevisiae}~\citep{Humphreys2021}, \textit{Escherichia coli}~\citep{Gao2022}, \textit{Bacillus subtilis}~\citep{OReilly2023}, and humans~\citep{Burke2023}. These structural models provide new insights into how proteins assemble into functional complexes responsible for processes ranging from transcription, translation and DNA repair to metabolism, biomolecular transport, and protein modification, and offer novel structural hypotheses for how interfacial mutations contribute to disease pathophysiologies.

\subsection{Designing binders}

\begin{figure}[htb!]
    \centering
    \includegraphics[width=213pt]{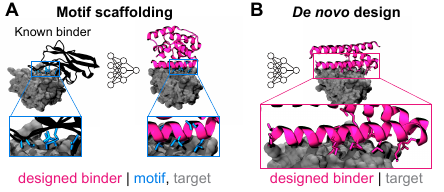}
    \caption{\textbf{Efforts to design binders using deep learning employ different strategies.} (A) One approach leverages knowledge of known (or predicted) binders of a given target protein and grafts their interaction motifs onto a scaffold to design a novel binder. (B) Another approach uses a generative model to \textit{de novo} design a binder, conditioned on its ability to bind the target. Because the design is not constrained to a particular interaction motif, novel interaction interfaces that maximize surface area and complementarity to the target can be potentially designed.}
    \label{fig:design_subtasks}
\end{figure}

Mechanistic knowledge of protein interactions ultimately enables the design of novel binders and ways to precisely modulate interaction networks. Computational methods to \textit{de novo} design proteins that specifically bind protein targets with high affinity have immense potential for engineering biosensors, therapeutics, synthetic biology tools, and biomaterials with broad basic and translational applications~\citep{Marchand2022,Lu2020}. One approach leverages knowledge of existing protein--protein interactions and their binding motifs to graft an existing interface onto a new protein scaffold, which is then optimized to ensure the structural integrity of the grafted binding motif and increase interaction affinity (Fig.~\ref{fig:design_subtasks}A)~\citep{Marchand2022}.
Deep learning methods for protein interaction and interaction site prediction can aid these efforts by expanding the pool of known interactions and guiding the initial choice of binding motif to seed designs, respectively~\citep{Gainza2023,Brixi2023}. For example, molecular surface fingerprint representations provided by the geometric deep learning framework MaSIF~\citep{Gainza2020,Sverrisson2020} have been used to construct a diverse set of initial designs by first identifying putative binding sites on a target protein and then efficiently selecting interacting motifs from a large database based on their fingerprint complementarity to the target site. Subsequent optimization, either experimentally or purely \textit{in silico}, yielded \textit{de novo} binders to the SARS-CoV-2 spike protein and immuno-oncology targets PD-1, PD-L1, and CTLA-4 with high affinity and/or specificity for the computationally designed interface~\citep{Gainza2023}.

However, such approaches have low success rates, with thousands of computational designs having to be experimentally screened to identify even a single low affinity binder in many cases~\citep{Gainza2023,Marchand2022}. This is largely attributed to designed sequences failing to either adopt the intended monomer structure or bind the intended surface~\citep{Bennett2023}. Since deep learning methods for structure prediction implicitly assess the probability that a sequence adopts some (potentially \textit{a priori} unknown) structure and forms a complex, these models can help mitigate these issues. One approach uses them to hallucinate novel binders: An initial amino acid sequence is iteratively updated to maximize the structure prediction model's confidence and optimize additionally desired properties, such as binding motif reconstruction, interaction quality, or complex symmetry~\citep{Wang2022_scaffolding_protein_functional_sites,Wicky2022,Goudy2023,Wang2023_EvoPlay,Jendrusch2021,Bryant2022_EvoBind,Yang2022,Hie2022}. Using deep network hallucination, experimentally validated protein and peptide binders~\citep{Wang2022_scaffolding_protein_functional_sites,Goudy2023,Wang2023_EvoPlay}, viral receptor traps~\citep{Wang2022_scaffolding_protein_functional_sites}, and giant oligomeric assemblies~\citep{Wicky2022} distinct from natural proteins have been designed. Integrating a deep learning model for structure-conditioned protein sequence generation, specifically ProteinMPNN~\citep{Dauparas2022}, into the optimization scheme further increases success rates by generating sequences with higher expression levels, solubilities, cyrstallization propensities, and thermostabilities~\citep{Wicky2022,Goudy2023}.

Most recently, deep generative models have shown even superior \textit{de novo} design capabilities (Fig.~\ref{fig:design_subtasks}B). By faithfully recapitulating the distribution of protein sequences or structures provided by training data, generative models enable facile sampling of realistic new proteins whose stability and expression matches that of natural ones. One especially successful class of generative model deployed presently is a diffusion model~\citep{Watson2023,Ingraham2023,Lin2023_genie}: Diffusion models transform complex data distributions into simple ones by gradually adding noise to training samples and learning to stochastically reverse this process in order to generate new data. By conditioning each step of reverse diffusion on specified design objectives, such as binding propensity, the model generates diverse, authentic proteins with desired properties from random noise. One protein diffusion model, RFdiffusion~\citep{Watson2023}, leverages the extensive knowledge of protein structure and complex assembly implicitly embedded in deep learning structure prediction models by repurposing the RoseTTAFold structure prediction network for reverse diffusion. Outperforming other \textit{de novo} design methods, RFdiffusion generates higher-order symmetric oligomers, protein scaffolds for binding motifs, and picomolar affinity peptide-binders without any subsequent optimization while requiring experimental testing of only tens of computational designs to identify successful ones~\citep{Watson2023,Vazquez2022}.

\section{Successes, limitations, and promising directions}

The challenge of understanding how molecularly diverse protein interactions collectively coordinate biological processes has spawned a growing suite of deep learning approaches (Fig.~\ref{fig:task_vs_type}). Of the methods reviewed here, deep neural networks for complex structure prediction~\citep{Jumper2021,Evans2022} and \textit{de novo} binder design~\citep{Watson2023,Vazquez2022} are preeminent examples of how synergistic use of data and biophysical priors yields highly accurate and general models~\citep{Ahdritz2023}. Based on the observation that physicochemical complementarity consistently distinguishes interacting proteins, structural information is increasingly leveraged to accomplish varied modeling tasks~\citep{Abdin2022,Jha2022,Krapp2023,Gainza2023,Gao2023}. Structure-based approaches accurately model stable, obligate interactions exhibiting limited conformation changes upon binding and under strong evolutionary pressures. However, they remain challenged by weak, transient interactions and protein interactors with extensive disordered regions that do not readily admit to experimental structure determination required for training and/or inference~\citep{Akdel2022,Yin2022,Burke2023,Lee2023,Gainza2020}. We anticipate that advancements across modeling tasks will continue to benefit from biologically-informed approaches that effectively utilize all available data and tailor network architectures and training regiments to suit different types of protein interactions.

\subsection{Expanding training datasets}

Depending on their biophysical properties, different classes of protein interactions are each amenable to different experimental characterization techniques. However, except for protein--protein interaction classifiers, the full breadth of available experimental data remains underutilized. Existing models for protein--peptide interaction prediction almost exclusively use training datasets derived from complex structures, limiting them to tens of thousands of interactions with well-defined bound structures~\citep{Lei2021,Palepu2022,Gurvich2023}. Retraining or fine-tuning them on the millions of qualitative binding measurements previously curated to construct (less performative) shallow machine learning models~\citep{Cunningham2020,AlQuraishi2014,Miller2008,Kundu2014} offers a straightforward route to improving them. Modern experimental methods for proteome-scale mapping of motif-mediated interactions~\citep{Benz2022,Davey2023} provide another, yet untapped, and continually growing source of data for training and evaluating models, with methods scaling to millions of peptides. A paucity of data has also stymied the development of deep learning models for interaction affinity prediction. This stands to change owing to recent experimental innovations that allow quantification of binding affinities at scale (tens of thousands of measurements in a day) with sensitivity for both strong (nanomolar) and weak (micromolar) interactions~\citep{Gogl2022}. These experimental methods provide data sorely needed to construct quantitative models that can account for differences in binding affinity that underlie interaction specificity, cooperativity, and competition; it is precisely these differences that enable systems of protein interactions to physically execute complex signaling logic and that explain how specific disease variants alter functional protein interaction networks~\citep{AlQuraishi2014}.

Efforts to build quantitative models of protein interactions may further benefit from approaches that harmonize small-scale, highly accurate measurements with large-scale, noisy experimental data to extract biophysically meaningful signals of binding that cover broad swaths of interaction space. Multitask learning provides one way to integrate heterogeneous data. For example, adding a classifier on top of AlphaFold2 and fine-tuning the combined network parameters for simultaneous binary classification of qualitative protein--peptide binding and complex structure prediction further improves the resulting model's discriminative capabilities on transient interactions~\citep{Motmaen2023}. Similar approaches could be devised to 
synthesize data from varied types of experimental assays~\citep{Yan2023}, ranging from \textit{in vitro} biophysical methods and those detecting complexes present in cell extracts or \textit{in vivo} to functional assays reporting on the cellular consequences of an interaction but not the physical interaction itself~\citep{Peng2017,Davey2023,Luck2020,Huttlin2021}. Directly accounting for known differences in readout, detection limits, sensitivities, and sources of uncertainty within the loss function~\citep{Lim2022}, for instance, or even explicitly modeling the steps of an experimental protocol~\citep{Rube2022} would provide a principled way to combine heterogeneous data that has been simplistically pooled to train many existing protein--protein interaction classifiers.

\subsection{Leveraging self-supervised representation learning}

Self-supervised representation learning offers another promising approach, pursuable in parallel with efforts to expand labeled training datasets, to improve the ability of deep learning models to reason about diverse protein interactions. For instance, sequence-based models using representations extracted from protein language models can match or even outperform structure-based models for predicting both protein interactions~\citep{Gurvich2023,Yang2023} and interaction sites~\citep{Wang2022_protein_peptide_binding_residues,Hosseini2023,Si2023_DRN-1D2D_Inter}. Crucially, they are applicable to proteins that have resisted experimental structure determination. One limitation of current protein language models, whose vocabularies comprise canonical amino acids, is their inability to account for post-translational modifications, which dynamically regulate protein interactions and whose misregulation is frequently associated with disease~\citep{Conibear2020}. In principle, post-translational modifications can be accounted for in structure-based models using atomistic representations. An emerging area of protein representation learning aims to better capture detailed spatiochemical information by directly learning from both sequences and structures~\citep{Zhang2023,Zhang2023_ESM-GearNet,Yang2023_MIF,Sun2023}. Such approaches have been used to distill geometric and physicochemical properties common among protein complex interfaces at the atomic level~\citep{MohseniBehbahani2023,Liu2021_GeoPPI}; however, more general-purpose representations provided by models trained on over an order-of-magnitude more monomer structures~\citep{Zhang2023,Zhang2023_ESM-GearNet} have yet to be incorporated within models of protein interactions. Furthermore, while current self-supervised approaches for learning representations of protein structure are quite promising, most directly borrow methods from natural language processing or computer vision~\citep{MohseniBehbahani2023,Zhang2023,Sun2023} and have not exploited domain-specific knowledge to craft learning objectives tailored to biomolecular structures~\citep{Liu2021_GeoPPI}.

\subsection{Incorporating biophysical priors}

Incorporating physical inductive biases can result in deep learning models that efficiently learn from limited data, better generalize to proteins outside their training distributions, and exhibit increased interpretability. For example, the geometric deep learning model for interaction site prediction \revision{ScanNet~\citep{Tubiana2022}} builds on the concept of pharmacophores to construct local, atomic-scale representations of amino acids that can be interpreted in terms of common chemical motifs, including hydrogen-bond networks and solvent-exposed basic residues; thus, the model provides testable hypotheses about molecular recognition mechanisms. The continued development of geometric deep learning architectures that account for inherent physical properties of molecular systems (\textit{e.g.,} SE(3)-equivariance) has yielded a number of successful approaches for encoding input monomer structures to model interactions in which proteins maintain `near-rigid' structures~\citep{Tubiana2022,Gainza2020,Sverrisson2020,Ganea2022,Krapp2023,Dai2021,Huang2023}. More nascent efforts have focused on developing neural network architectures that effectively utilize monomer structures while simultaneously allowing for conformational changes upon binding~\citep{Chu2023,McPartlon2023,Abdin2022}. Such architectures may facilitate the development of methods that enable explicit encoding of conformational ensembles, perhaps through generative models. This promises advancements in modeling interactions that involve substantial structural changes and interactions mediated by intrinsically disordered proteins~\citep{Janson2023,Tesei2023}, both of which remain challenging for current methods. Key to training deep learning models that predict equilibrium conformational ensembles has been the use of data from physics-based molecular simulations~\citep{Noe2019,Zheng2023_DiG,Tesei2023}, which resolve equilibrium ensembles with substantially higher resolution and efficiency than current experimental approaches.

More broadly, physics-informed machine learning encompasses a variety of approaches that integrate experimental observations with physical models to facilitate learning of solutions that respect underlying physical constraints (\textit{e.g.,} symmetries, conservation laws, thermodynamic equilibrium, monotonicity)~\citep{Karniadakis2021}. Such methods promise to be valuable for building predictors of protein interaction thermodynamics and kinetics. One approach, complementary to strategies described previously, incorporates known physical functional forms to learn binding energy functions in which intermolecular interactions scale with distance, as expected from quantum mechanical arguments~\citep{Sverrisson2022}. Another learns approximate steady-state solutions to ordinary differential equations customarily employed within dynamical models of interaction networks that saturate at physiological concentrations~\citep{Nilsson2022}. One flexible framework for learning models that approximately satisfy generic physical constraints introduces biases through the choice of optimization objectives and inference algorithms. This has been used to learn generative models of protein conformational ensembles that recapitulate the Boltzmann distribution~\citep{Noe2019,Zheng2023_DiG} and neural network time-series models of transition path ensembles linking different protein conformational states~\citep{Tsai2022}. Applying such approaches to protein interactions, and not just individual proteins, may provide insights into binding mechanisms.

Integrated modeling frameworks also provide a promising path to expand models beyond individual protein complexes to ultimately encompass entire networks embedded with their cellular context. Formulating deep learning models in terms of an appropriate statistical mechanical ensemble~\citep{Cunningham2020,AlQuraishi2014}, standard practice in physics-based modeling, provides a natural way to account for variations in physical conditions (\textit{e.g.,} temperature, pH) or protein abundances characteristic of different environmental conditions or cellular states. Predicting how proteome-scale interaction networks coordinate cellular responses to environmental changes or therapeutic interventions additionally serves to benefit from approaches that integrate deep learning techniques with existing dynamical systems models~\citep{Yuan2021,Stapor2022} or that directly exploit mathematical correspondences between the two types of models~\citep{Chen2018,Rackauckas2021}. In the quest for a comprehensive understanding of protein interaction networks and their roles in complex cellular behaviors, a variety of innovations will likely be needed. One aspect nevertheless seems certain: Given their demonstrated value for discovering novel interactions, elucidating molecular recognition mechanisms, and designing functional assemblies, deep learning methods will remain a flourishing component of the larger ecosystem of quantitative biology models of protein interactions.

\begin{appendices}
\section{Appendix: Ecosystem of deep learning methods for modeling protein interactions}

\begin{itemize}
\item \textbf{(1.i) Predictors of direct protein--peptide interactions:} Peptriever~\citep{Gurvich2023}

\item \textbf{(1.ii) Predictors of direct protein--protein interactions:} \cite{Unsal2022}

\item \textbf{(1.iii) Predictors of higher-order complexes:} PCGAN~\citep{Pan2023}

\item \textbf{(1.ii-iii) Predictors of direct protein--protein interactions and higher-order complexes:} DeepPPI~\citep{Du2017}, DeepSequencePPI~\citep{Gonzalez-Lopez2018}, DPPI~\citep{Hashemifar2018}, DNN-PPI~\citep{Li2018}, PIPR~\citep{Chen2019}, \cite{Richoux2019}, DeepFE-PPI~\citep{Yao2019}, Multimodal SAE~\citep{Zhang2019_multimodal_deep_representation_learning}, EnsDNN~\citep{Zhang2019_EnsDNN}, \cite{Jha2020}, EnAmDNN~\citep{Li2020_EnAmDNN}, \cite{Liu2020_binary_ppi_prediction}, \cite{Yang2020}, S-VGAE~\citep{Yang2020_SVGAE}, AutoPPI~\citep{Czibula2021_AutoPPI}, MTT~\citep{Dong2021}, DeepViral~\citep{Liu-Wei2021}, FSNN-LGBM~\citep{Mahapatra2021}, FSFDW~\citep{Nasiri2021}, D-SCRIPT~\citep{Sledzieski2021}, S2F~\citep{Xue2021_S2F}, TransPPI~\citep{Yang2021}, DeepTrio~\citep{Hu2022_DeepTrio}, \cite{Jha2022}, SDNN-PPI~\citep{Li2022_SDNN-PPI}, Topsy-Turvy~\citep{Singh2022}, TAGPPI~\citep{Song2022}, RAPPPID~\citep{Szymborski2022}, P-PPI~\citep{Anteghini2023}, HIGH-PPI~\citep{Gao2023}, SYNTERACT~\citep{Hallee2023}, SGPPI~\citep{Huang2023}, \cite{Jha2023_analyzing_effect_of_multi_modality} CNN, \cite{Jha2023_PPI-Graph-BERT} GNN, SENSE-PPI~\citep{Volzhenin2023}, PPITrans~\citep{Yang2023}, SemiGNN-PPI~\citep{Zhao2023}

\item \textbf{(1-2.i) Predictors of protein--peptide interactions and interaction sites:} CAMP~\citep{Lei2021}, Cut\&CLIP~\citep{Palepu2022}

\item \textbf{(1-2.ii) Predictors of direct protein--protein interactions and interaction sites:} MaSIF~\citep{Gainza2020}, dMaSIF~\citep{Sverrisson2020}

\item \textbf{(2.i) Predictors of protein--peptide interaction sites:} Visual \citep{Wardah2020}, BiteNetP${_{\rm P_P}}$~\citep{Kozlovskii2021}, PepNN~\citep{Abdin2022}, PepBCL~\citep{Wang2022_protein_peptide_binding_residues}, PepCNN~\citep{Chandra2023}, GraphPPepIS and SeqPPepIS~\citep{Li2023}

\item \textbf{(2.ii) Predictors of direct protein--protein interaction sites:} \cite{Fout2017}, ComplexContact \citep{Zeng2018}, SASNet~\citep{Townshend2019}, DLPred~\citep{Zhang2019_DLPred}, \cite{Liu2020}, TopNetTree~\citep{Wang2020_TopNetTree}, \cite{Xie2020}, MuPIPR~\citep{Zhao2020}, PInet~\citep{Dai2021}, DELPHI~\citep{Li2021}, GeoPPI~\citep{Liu2021_GeoPPI}, DNCON2\_Inter~\citep{Quadir2021}, DeepHomo~\citep{Yan2021}, DLA~\citep{MohseniBehbahani2023}, PITHIA~\citep{Hosseini2022}, DRCon~\citep{Roy2022}, PIPENN~\citep{Stringer2022}, ScanNet~\citep{Tubiana2022}, PGT~\citep{Wu2022}, GLINTER~\citep{Xie2022}, SaLT\&PepPer~\citep{Brixi2023} Seq-InSite~\citep{Hosseini2023}, EDLMPPI~\citep{Hou2023}, PeSTo~\citep{Krapp2023}, ISPRED-SEQ~\citep{Manfredi2023}, DRN-1D2D\_Inter~\citep{Si2023_DRN-1D2D_Inter}, PLMGraph-Inter~\citep{Si2023_PLMGraph-Inter}

\item \textbf{(3.ii) Predictors of direct protein--protein complex structures (and interaction sites):} ESMPair~\citep{Chen2022_ESMPair}, EquiDock~\citep{Ganea2022}, \cite{Sverrisson2022}, GeoDock~\citep{Chu2023}, RoseTTAFold2~\citep{Baek2023}, AFProfile~\citep{Bryant2023_AFProfile}, AlphaRED~\citep{Harmalkar2023}, DiffDock-PP~\citep{Ketata2023}, AFsample~\citep{Wallner2023}

\item \textbf{(3.i-ii) Predictors of protein--peptide and direct protein--protein complex structures (and interaction sites):} AF-linker~\citep{Tsaban2022,Ko2021}, \cite{Lee2023}

\item \textbf{(1-3.i) Predictors of protein--peptide interactions and complex structures (and interaction sites):} \cite{Motmaen2023}, \cite{Chang2023}, \cite{Teufel2023}

\item \textbf{(3.ii-iii) Predictors of direct protein--protein interactions and higher-order complexes structures (and interaction sites):} AF2Complex~\citep{Gao2022}, AlphaLink~\citep{Stahl2023,Stahl2023_AFLink-multimer}, ESMFold-linker~\citep{Lin2023}

\item \textbf{(3.iii) Predictors of higher-order complex structures:} MolPC \citep{Bryant2022_MolPC}, Uni-Fold Symmetry~\citep{Li2022_UniFold_Symmetry}, ~\cite{Jeppesen2023}, CombFold~\citep{Shor2023}

\item \textbf{(1-3.i-iii) Predictors of protein--peptide, direct protein--protein, and higher-order complex interactions, interaction sites, and complex structures:} RosseTTAFold-gap~\citep{Baek2021,Tsaban2022}, AF2-gap~\citep{Bryant2022_FoldDock}, AF2-Multimer~\citep{Evans2022}, ColabFold~\citep{Mirdita2022}, AlphaPulldown~\citep{Yu2023}, OmegaFold-linker~\citep{Tsuchiya2022,Shanker2023}

\item \textbf{(3-4.ii) Predictors of direct protein--protein complex structures (and interaction sites) and models to design protein binders:} DockGPT~\citep{McPartlon2023}

\item \textbf{(4.i) Models to design peptide binders:} EvoBind~\citep{Bryant2022_EvoBind}, GANDALF~\citep{Rossetto2020}, \cite{Yang2022}, PepPrCLIP~\citep{Bhat2023}

\item \textbf{(4.ii) Models to design protein binders:} \cite{Gainza2023}, EvoPro~\citep{Goudy2023}, ProteinEnT~\citep{Mahajan2023}, EvoPlay~\citep{Wang2023_EvoPlay}

\item \textbf{(4.iii) Models to design higher-order complexes:} Chroma \citep{Ingraham2023}

\item \textbf{(4.i-ii) Models to design peptide and protein binders:} RF\textsubscript{joint}~\citep{Wang2022_scaffolding_protein_functional_sites}

\item \textbf{(4.ii-iii) Models to design protein binders and higher-order complexes:} AlphaDesign~\citep{Jendrusch2021}, \cite{Hie2022}, ESM-IF1~\citep{Hsu2022}

\item \textbf{(4.i-iii) Models to design peptide and protein binders and higher-order complexes:} Hallucination with structure prediction models (RoseTTAFold, trRosetta~\citep{Wang2022_scaffolding_protein_functional_sites}, or AF2~\citep{Wicky2022}), RFDiffusion~\citep{Watson2023,Vazquez2022}

\end{itemize}

\end{appendices}

\section{Competing interests}
M.A. is a member of the Scientific Advisory Boards of Cyrus Biotechnology, Deep Forest Sciences, Nabla Bio, Oracle Therapeutics, and FL2021-002, a Foresite Labs company.

\section{Author contributions statement}
J.R.R., G.N., and M.A. conceived, designed, wrote, and edited the manuscript.

\section{Acknowledgments}
J.R.R. acknowledges a Fellowship of The Jane Coffin Childs Memorial Fund for Medical Research. We thank Eric Wong, Tony Culos, Martin Culka, Christina Floristean, Lukas Jarosch, Yeqing Lin, Harry Lee, Pavan Ravindra, Charlotte Rochereau, Toni Sagayaraj, Jae Shin, Wojtek Treyde, Arthur Valentin, and the rest of the AlQuraishi group for valuable discussions about modeling biomolecules using deep learning.

\bibliographystyle{abbrvnat_peds}
\bibliography{refs}


\end{document}